\documentclass[12pt]{iopart}
\usepackage[colorlinks=true,citecolor=blue,linkcolor=blue,urlcolor=blue]{hyperref}
\usepackage{xcolor}
\usepackage{graphicx}
\usepackage{amssymb}
\usepackage{bbm}
\usepackage{iopams}
\usepackage{cite}

\begin{document}

\title{Optical nonreciprocity induced by quantum squeezing in temperature sensitive optomechanical  systems}

\author{Jun-Cong Zheng, Xiao-Wei Zheng, Xin-Lei Hei, Yi-Fan Qiao, Xiao-Yu Yao, Xue-Feng Pan, Yu-Meng Ren, Xiao-Wen Huo and Peng-Bo Li$^{*}$}

\address{Ministry of Education Key Laboratory for Nonequilibrium Synthesis and Modulation of Condensed Matter, Shaanxi Province Key Laboratory of Quantum Information and Quantum Optoelectronic Devices, School of Physics, Xi’an Jiaotong University, Xi’an 710049, China}
\vspace{10pt}

\begin{indented}
	\item[\textbf{E-mail:}]~lipengbo@mail.xjtu.edu.cn
\end{indented}

\begin{indented}
	\item[\textbf{Keywords:}]~nonreciprocal, quantum squeezing, optomechanical systems
\end{indented}

\begin{abstract}
We investigate single-photon transmission and the statistical properties of photon correlations in $\chi^{(2)}$ microring optomechanical systems, where optical nonreciprocity is induced by directional quantum squeezing. Due to the presence of thermal phonons in the mechanical resonator, the system is highly sensitive to temperature changes.
Our numerical simulations show that as the thermal phonons vary from 0 to 10, the isolation ratio of single-photon transmission decreases from 22.2 dB to 1.1 dB (or from -23 dB to -3.3 dB). Additionally, the statistical properties of photon correlations transition from exhibiting a strong bunching effect to a weak bunching effect. Moreover, the parametric amplification component enhances the device's temperature response, distinguishing it from other similar nonreciprocal devices.
Our protocol suggests a potential application for nonreciprocal setups in precise temperature measurement at ultralow temperatures, thereby enriching quantum networks and quantum information processing.
\end{abstract}

\section{Introduction}\label{I}

Optomechanical systems, which explore the interaction between electromagnetic radiation and nanomechanical elements, have advanced significantly in recent years \cite{kippenberg2008cavity,PhysRevLett.109.063601,PhysRevLett.110.253601,RevModPhys.86.1391,xu2016topological,PhysRevLett.129.063605,PhysRevLett.132.053601}. Based on the radiation modes and vibrational degrees of freedom, optomechanical systems can be categorized into different types of setups, including suspended micro-pillars \cite{jansen2023nanocrystal}, suspended membranes \cite{woolf2013optomechanical}, optical microsphere resonators \cite{PhysRevA.87.055802}, and near-field coupled nanomechanical oscillators \cite{anetsberger2009near}. The last type of setup is commonly found in microring optomechanical systems, which benefit from directional evanescent coupling \cite{PhysRevApplied.5.054019,lodahl2017chiral,PhysRevLett.111.193601,PhysRevLett.110.213604,PhysRevA.90.043802,PhysRevA.99.043833,zheng2023few}. In these microring optomechanical systems, there are both clockwise (CW) and counterclockwise (CCW) modes in the optical resonator, making them suitable for achieving optical nonreciprocity by inducing the splitting of degenerate modes \cite{hafezi2012optomechanically,shen2018reconfigurable,shen2016experimental,ruesink2016nonreciprocity,ruesink2018optical,xu2020quantum}. Compared to traditional purely optical nonreciprocal devices \cite{PhysRevApplied.10.047001,nassar2020nonreciprocity,PhysRevLett.124.070402,kutsaev2021up,PhysRevA.105.013711,PRXQuantum.4.010306,PhysRevA.109.063709}, these optomechanical systems can link nonreciprocal effects to ambient temperature, offering potential applications in temperature measurement at ultralow temperatures \cite{xu2020quantum}. However, weak optomechanical coupling remains a challenge for achieving precise measurements, and strengthening the optomechanical coupling in these nonreciprocal systems is a key focus for future research.

Quantum squeezing \cite{esteve2008squeezing,drummond2013quantum,ma2011quantum,PhysRevA.71.055801}, an important resource in the field of quantum information, is typically generated through fully quantum degenerate parametric amplifiers (DPAs) \cite{PhysRevLett.127.093602,PhysRevLett.129.123602} or semiclassical DPAs \cite{PhysRevLett.120.093601,PhysRevLett.126.023602,PhysRevA.100.062501}. It is often applied to manipulate zero-point motion, enabling the realization of ultrasensitive sensing of force and motion \cite{barnett1987squeezing,PhysRevA.46.R6797,wollman2015quantum,PhysRevLett.115.243601,PhysRevLett.117.100801,lawrie2019quantum,xu2019sensing,PhysRevX.13.041021,lu2023recent}. Recent research has focused on quantum squeezing in a $\chi^{(2)}$ resonator mode, achieving exponentially enhanced optomechanical coupling \cite{PhysRevLett.114.093602}. Other studies have realized optical nonreciprocity through directional quantum squeezing in  $\chi^{(2)}$ microring optical systems \cite{PhysRevLett.128.083604,PhysRevA.108.023716}. Building on these contributions, precise temperature measurement using nonreciprocal devices, as mentioned earlier, becomes feasible.

In this work, we simultaneously achieved optical nonreciprocity and enhanced optomechanical coupling, both induced by quantum squeezing in a microring optomechanical system. Specifically, we incorporate an additional mechanical resonator coupled to the optical resonators, building on the configuration in Refs. \cite{PhysRevLett.128.083604,PhysRevA.108.023716}. In this case, the transmission of incident photons is influenced by thermal phonons. Our numerical simulations show that as the temperature changes, the effectiveness of single-photon nonreciprocal transmission diminishes, and the statistical properties of photons evolve from exhibiting a strong bunching effect to a weak bunching effect. Additionally, quantum squeezing enhances the optomechanical coupling between the optical resonator and the mechanical resonator. As show in Fig. \ref{fig.1}, we placed the oscillator in three different regions, including regions with reinforced optomechanical coupling and regions where the coupling remains invariant. The results indicate that the directional parametric amplification component (with the mechanical oscillator placed in Area I or Area III) has a more pronounced effect on photon transmission. In conclusion, our method provides a platform for achieving optical nonreciprocity in temperature-sensitive optomechanical systems. This approach could potentially be extended to acoustic modes, where nonlinear acoustic devices can be realized by coupling to a two-level system \cite{PhysRevA.80.033846,PhysRevA.82.032101}. Acoustic diodes, which are widely used in phonon-based quantum information processing \cite{PhysRevLett.103.104301,maznev2013reciprocity,PhysRevLett.106.084301,PhysRevB.99.214305,PhysRevApplied.3.064014}, could benefit from such advancements.

This work is structured as follows: In Section \ref{II}, we introduce the model of temperature-sensitive optomechanical systems and present the corresponding Hamiltonian. Moving on to Sections \ref{III} and \ref{IV}, we show the response of single-photon nonreciprocal transport and the statistical properties of correlated photons to temperature changes, respectively. Section \ref{V} is dedicated to a discussion on experimental feasibility, and finally, Section \ref{VI} provides a summary of the findings.

\section{Physical model}\label{II}

\begin{figure}
	\centering
	\includegraphics[scale=0.6]{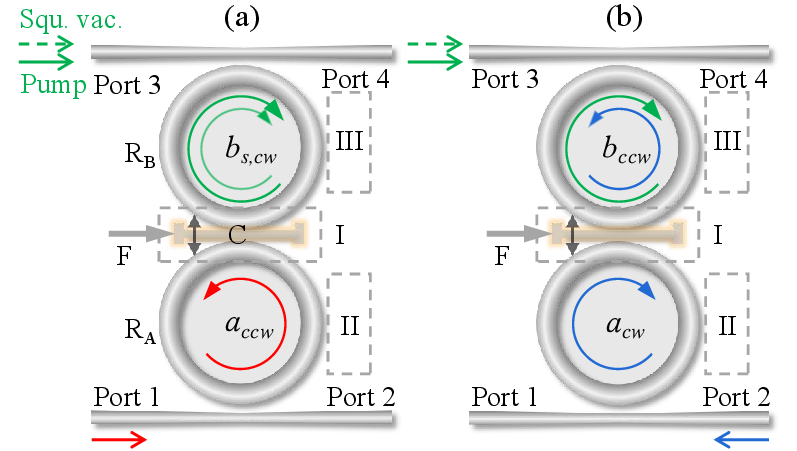}
	\caption{Schematic diagram illustrating the generation of ambient temperature-sensitive nonreciprocal single-photon isolation and nonreciprocal photon correlations. Two microring resonators, named $R_A$ and $R_B$, are coupled to a lower bus waveguide (at a rate $\kappa_a$) and an upper drop waveguide (at a rate $\kappa_b$), respectively. The mechanical nanostring oscillator $C$ (driven by a force $F$) is placed at three different locations: areas I, II, and III. A coherent pump field and a broadband squeezed-vacuum field are input from port 3, with the squeezed CW mode in $R_B$ denoted as $b_{s,cw}$. (a) A probe field input from port 1 excites a CCW mode $a_{ccw}$ in $R_A$, which interacts with the oscillator $C$ and the mode $b_{s,cw}$, respectively. (b) A probe field input from port 2 excites a CW mode $a_{cw}$ in $R_A$, which interacts with the oscillator $C$ and the mode $b_{ccw}$, respectively.}
	\label{fig.1}
\end{figure}

We aim to correlate the nonreciprocal effect induced by quantum squeezing with ambient temperature in a microring optomechanical system. By leveraging the strong optomechanical coupling generated through parametric amplification \cite{PhysRevLett.114.093602}, this setup shows potential for applying nonreciprocal systems in precise temperature measurements.
The schematic of the proposed system is illustrated in Fig. \ref{fig.1}. Two microring resonators, $R_A$ and $R_B$, with $\chi^{(2)}$ nonlinearity, are made of high-quality thin film and interact with each other via the optical evanescent field with coupling strength $g_0$. The mechanical resonator $C$ is placed in area I, located in the middle of the gap between $R_A$ and $R_B$, and is simultaneously optomechanically coupled to these two optical resonators with coupling strength $j_0$ but with opposite signs \cite{xu2020quantum, agasti2023back} (refer to \hyperref[AppendixA]{Appendix A} for details). In this case, the semi-classical optomechanical Hamiltonian of the mechanical mode is expressed as (setting $\hbar=1$)

\begin{equation}\label{eq1}
H_m^{\scriptstyle \mathrm{I}}=\frac{1}{2}
\omega_m(q^2+p^2)-\sum_{\xi=cw,ccw}j_0 (a^{\dagger}_{\xi}a_{\xi}-b^{\dagger}_{\xi}b_{\xi})q,
\end{equation}
where $a_{\xi}$, $b_{\xi}$, $a_{\xi}^{\dagger}$, and $b_{\xi}^{\dagger}$ ($\xi=cw, ccw$) denote the annihilation and creation operators of optical modes with frequencies $\omega_a$ and $\omega_b$, respectively, $q$ and $p$ represent the dimensionless displacement and momentum operators of the mechanical resonator with frequency $\omega_m$. These operators for the mechanical resonator can be expressed in terms of phonon creation and annihilation operators as $q=(c^{\dagger}+c)/\sqrt{2}$ and $p=i(c^{\dagger}-c)/\sqrt{2}$. Consequently, the fully quantized Hamiltonian is given by:
\begin{equation}\label{eq2}
\mathcal{H}_m^{\scriptstyle \mathrm{I}}=\omega_m c^{\dagger}c-\sum_{\xi=cw,ccw}J_0(a^{\dagger}_{\xi}a_{\xi}-b^{\dagger}_{\xi}b_{\xi})(c^{\dagger}+c),
\end{equation} 
where $J_0 = j_0/\sqrt{2}$, the constant term $\omega_m/2$ can be nullified by selecting an appropriate gauge. Additionally, the resonator $C$, positioned in areas II and III, interacts with the single resonators $R_A$ and $R_B$, respectively. The corresponding Hamiltonians are expressed as:

\begin{equation}
\mathcal{H}_m^{\scriptstyle \mathrm{II}}=\omega_m c^{\dagger}c-\sum_{\xi=cw,ccw}J_0 a^{\dagger}_{\xi}a_{\xi}(c^{\dagger}+c),
\end{equation}\label{eq3} 

\begin{equation}
\mathcal{H}_m^{\scriptstyle \mathrm{III}}=\omega_m c^{\dagger}c-\sum_{\xi=cw,ccw}J_0b^{\dagger}_{\xi}b_{\xi}(c^{\dagger}+c).
\end{equation}\label{eq4}

Now, we discuss how the configuration works. We consider that $C$ is initially placed in area I (corresponding to the Hamiltonian $\mathcal{H}_m^{\scriptstyle \mathrm{I}}$). A continuous-wave coherent laser field with frequency $\omega_s$ and strength $\Omega$, pumped from port 3, squeezes the CW mode $b_{cw}$ in $R_B$ into $b_{s,cw}$ under the two-mode phase-matching condition \cite{PhysRevLett.128.083604,PhysRevA.108.023716, PhysRevLett.126.133601}, while remaining decoupled from the CCW mode $b_{ccw}$. As shown in Fig. \ref{fig.1}(a), a weak probe field with frequency $\omega_p$ and amplitude $\varepsilon$, input from port 1, excites the CCW mode $a_{ccw}$ in $R_A$.
Due to the pump field, mode $a_{ccw}$ interacts with the squeezed mode $b_{s,cw}$ at an effective rate $g_s$. Regarding the optomechanical coupling, the oscillator \( C \) interacts with the squeezed mode \( b_{s,cw} \) at an effective rate \( J_s \), and with the mode \( a_{ccw} \) at the unmodulated rate \( J_0 \). It is important to note that we neglect the coupling of oscillator \( C \) with the CCW mode in \( R_B \) and the CW mode in \( R_A \) in the following discussion, as their effects on photon transmission in the system are negligible (refer to \hyperref[AppendixC]{Appendix C} for details).

In the rotating reference frame defined by the unitary operator $R(t) = \textrm{exp}[i\omega_s(a^{\dagger}_{ccw}a_{ccw} + b^{\dagger}_{cw}b_{cw})t/2]$, the Hamiltonian of the entire system for the case of a probe field input from port 1 is expressed as:

\begin{eqnarray}\label{eq5} 
\mathcal{H}_1&=\mathcal{H}_0+\mathcal{H}_{int}+\mathcal{H}_p+\mathcal{H}_s+\mathcal{H}_f, \nonumber\\
\mathcal{H}_0&=\Delta_s^a a^{\dagger}_{ccw}a_{ccw}+\Delta_s^b b^{\dagger}_{cw}b_{cw}+\omega_m c^{\dagger}c,    \nonumber\\
\mathcal{H}_{int}&=g_0(a^{\dagger}_{ccw}b_{cw}+a_{ccw}b^{\dagger}_{cw})-J_0(a^{\dagger}_{ccw}a_{ccw}-b^{\dagger}_{cw}b_{cw})(c^{\dagger}+c),    \nonumber\\ 
\mathcal{H}_p&=i\sqrt{\kappa_{ex1}}(\varepsilon a^{\dagger}_{ccw}e^{-i\Delta_p t}-\varepsilon^* a_{ccw}e^{i\Delta_p t}), \nonumber\\
\mathcal{H}_s&=\frac{\Omega}{2}(b^{\dagger 2}_{cw}+b^2_{cw}),\nonumber\\ \mathcal{H}_f&=F(c^{\dagger}+c),
\end{eqnarray}
where $\mathcal{H}_0$ represents the free Hamiltonian for the microring resonators and the mechanical oscillator. $\mathcal{H}_{int}$ denotes the interaction Hamiltonian, with the first term representing pure photon exchange and the second term reflecting the exchange between photons and phonons. $\mathcal{H}_p$ and $\mathcal{H}_s$ denote the Hamiltonians of the probe field and the squeezing pump field, respectively. Additionally, we introduce a constant Hamiltonian $\mathcal{H}_f$ for the mechanical mode to counteract the force induced by the parametric amplification  (see below).
The detunings are $\Delta_s^{a/b}=\omega_{a/b}-\omega_s/2$, $\Delta_p=\omega_p-\omega_s/2$, and $\kappa_{ex1}$ is the external decay rate of resonator $R_A$. 

To calculate the scattering of incident photons from port 1, we can transform the Hamiltonian $\mathcal{H}_1$ into the squeezing picture. By applying the Bogoliubov squeezing transformation $b_s=\textrm{cosh}(r_s)b+\textrm{sinh}(r_s)b^{\dagger}$ \cite{PhysRevLett.127.093602,PhysRevLett.114.093602}, with the squeezing parameter $r_s=\frac{1}{4}\textrm{ln}[(1+\beta)/(1-\beta)]$, where $\beta=\Omega/\Delta_s^b$ is the pump ratio, the Hamiltonian in the frame rotating at the frequency of $\Delta_p$ becomes (refer to \hyperref[AppendixB]{Appendix B} for details):

\begin{eqnarray}\label{eq6} 
	\mathcal{H}^s_1&=&\Delta_a a^{\dagger}_{ccw}a_{ccw}+i\sqrt{\kappa_{ex1}}(\varepsilon a^{\dagger}_{ccw}-\varepsilon^* a_{ccw})+\Delta_m c^{\dagger}c\nonumber \\&&+\Delta_b^s b^{\dagger}_{s,cw}b_{s,cw}+g_s(a^{\dagger}_{ccw}b_{s,cw}+a_{ccw}b^{\dagger}_{s,cw})\nonumber \\&&-(J_0 a^{\dagger}_{ccw}a_{ccw}-J_s b^{\dagger}_{s,cw}b_{s,cw})(c^{\dagger}+c).
\end{eqnarray}
We have neglected the parametric interaction \cite{PhysRevLett.114.093602} and the counter-rotating terms under the rotating-wave approximation ($g_0 \textrm{sinh}(r_s) \ll \Delta_s^a + \Delta_s^b \sqrt{1 - \beta^2}$). Additionally, the induced force applied to the mechanical oscillator is canceled by the additional force $F =- J_0 \textrm{sinh}^2(r_s)$, which remains constant once the squeezing parameter is fixed. 
We define $\Delta_a=\Delta_s^a-\Delta_p$, $\Delta_b^s=\Delta_s^{bs}-\Delta_p$, $\Delta_s^{bs}=\Delta_s^{b}\sqrt{1-\beta^2}$, $\Delta_m=\omega_m-\Delta_p$, and $g_s=g_0\textrm{cosh}(r_s)$, $J_s=J_0\textrm{cosh}(2 r_s)$. Here, $\Delta_b^s$ and $g_s$, $J_s$ are the effective squeezed mode detuning and the effective couplings, respectively, adjusted by the squeezing pump field.

\begin{figure}
	\centering
	\includegraphics[scale=0.55]{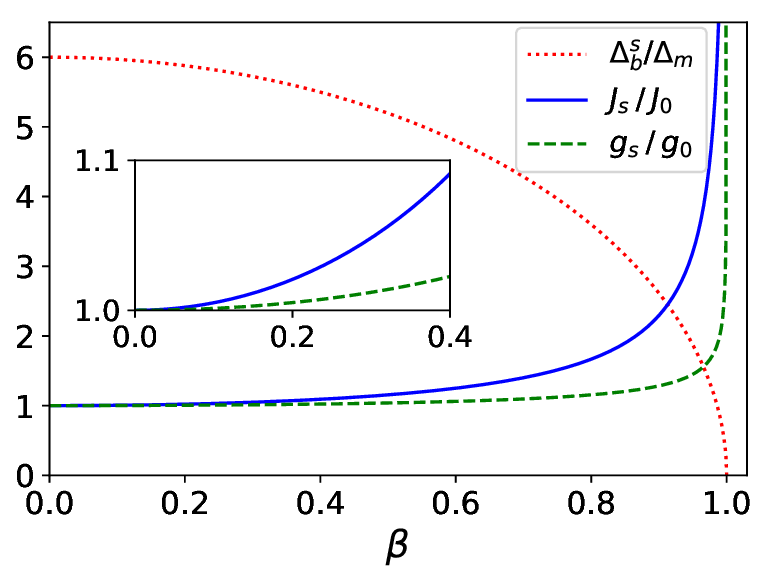}
	\caption{The squeezing parameters, including the effective squeezed mode detuning $\Delta_b^s$ (red dotted line), the effective coupling rates $J_s$ (blue solid line), and $g_s$ (green dashed line), are shown as functions of the pump ratio $\beta \in [0,1)$. Other parameters are set as $\Delta_s^b = 6\Delta_m$ and $\Delta_p = 0$.}
	\label{fig.2}
\end{figure}

To quantify the influence of the squeezing pump field $\mathcal{H}_s$, we plot the detuning $\Delta_b^s/\Delta_m$ and the coupling ratios $J_s/J_0$ and $g_s/g_0$ versus the pump ratio $\beta$ in Fig.\ref{fig.2}. Initially, at $\beta=0$, the parameters correspond to the normal mode situation. As $\beta$ increases, $\Delta_b^s/\Delta_m$ exhibits a parabolic decrease, eventually approaching 0 near $\beta=1$. For the coupling ratios, before $\beta=0.9$, $J_s/J_0$ and $g_s/g_0$ increase gradually, with $g_s/g_0$ increasing slightly more slowly than $J_s/J_0$, indicating that the coupling with the mechanical nanostring oscillator is more responsive to the squeezing pump. Additionally, the simulation shows that both the squeezed detuning and the coupling react dramatically (i.e., exponentially decreasing and exponentially increasing, respectively) as the pump ratio approaches $\beta=1$.
From a physical perspective, the enhancement in coupling strength arises from a single-photon state in the squeezed mode, which corresponds to an exponentially increasing number of photons in the $R_B$ as the squeezing strength increases. Consequently, the radiation pressure exerted by a single squeezed photon on both the $R_A$ and the mechanical resonator also increases, thereby strengthening the coupling with the squeezed cavity mode. Furthermore, the optical coupling grows much more slowly than the optomechanical coupling, as the former’s coupling strength is primarily determined by the amplitude of the optical field and does not directly depend on changes in photon number. In contrast, the latter’s coupling strength is directly dependent on the photon number within the squeezed cavity.

In Fig.\ref{fig.1}(b), the weak probe field input from port 2 excites the CW mode $a_{cw}$, which interacts with the CCW mode $b_{ccw}$ in $R_B$ at a rate of $g_0$. Regarding the optomechanical coupling, the oscillator $C$ interacts with both mode $a_{cw}$ and mode $b_{ccw}$ at a rate of $J_0$.  we neglect the coupling of oscillator \( C \) with the CW mode in \( R_B \) and the CCW mode in \( R_A \) in the following discussion, as their effects on photon transmission in the system are negligible (refer to \hyperref[AppendixC]{Appendix C} for details).
In this case, the squeezing pump field can be approximated as having no impact on the scattering of the incident photon, and the Hamiltonian is given by:

\begin{eqnarray}\label{eq7}  
	\mathcal{H}_2&=&\Delta_a a^{\dagger}_{cw}a_{cw}+i\sqrt{\kappa_{ex1}}(\varepsilon a^{\dagger}_{cw}-\varepsilon^* a_{cw})+\Delta_m c^{\dagger}c\nonumber \\&&+\Delta_b b^{\dagger}_{ccw}b_{ccw}+g_0(a^{\dagger}_{cw}b_{ccw}+a_{cw}b^{\dagger}_{ccw})\nonumber \\&&-J_0(a^{\dagger}_{cw}a_{cw}-b^{\dagger}_{ccw}b_{ccw})(c^{\dagger}+c),
\end{eqnarray}
where $\Delta_b = \Delta_s^b - \Delta_p$. Comparing the Hamiltonians $\mathcal{H}^s_1$ and $\mathcal{H}_2$, due to directional quantum squeezing, the effective detuning of resonator $R_B$ and the couplings are significantly different. If $\Delta_s^b$ is set to a small value, the effect of quantum squeezing on the detuning decreases, and the primary difference lies in the couplings.

In this work, we focus on single-photon transmission and the statistical properties of correlated photons. In the squeezing picture, the evolution of the entire system without squeezed-vacuum driving, as shown in Fig.\ref{fig.1}(a), can be characterized by the following master equation:

\begin{eqnarray}\label{eq8} 
	\frac{d \rho_1}{d t}&=&-i[\mathcal{H}_1^s,\rho_1]+\kappa_a L[a_{ccw}]\rho_1+\kappa_b L[b_{s,cw}]\rho_1\nonumber \\&&+\gamma_m(n_{\scriptstyle \mathrm{th}}+1)L[c]\rho_1+\gamma_m n_{\scriptstyle \mathrm{th}}L[c^{\dagger}]\rho_1\nonumber \\&&+\mathcal{L}_n[b_{s,cw}]\rho_1,
\end{eqnarray}
where $L[o]\rho = 2o\rho o^{\dagger} - o^{\dagger}o\rho - \rho o^{\dagger}o$ is the Lindblad superoperator for operator $o$, and $\rho_1$ is the system density matrix. The total decay rates of resonators $R_A$ and $R_B$ are $\kappa_a = \kappa_{ex1} + \kappa_i$ and $\kappa_b = \kappa_{ex2} + \kappa_i$, respectively. Here, $\kappa_i$ is the intrinsic decay rate of the optical resonators, and $\kappa_{ex2}$ is the external decay rate of resonator $R_B$. In the following, we assume $\kappa_a = \kappa_b = \kappa$ and $\Delta_a = \Delta_b = \Delta$ for simplicity.
The decay rate of the mechanical resonator $C$ is $\gamma_m$, and the mean thermal phonon number $n_{\scriptstyle \mathrm{th}} = 1/[\textrm{exp}(\omega_m / k_B T) - 1]$, where $k_B$ is the Boltzmann constant and $T$ is the temperature of the reservoir at thermal equilibrium. The last term, $\mathcal{L}_n[b_{s,cw}]\rho_1$, describes the effective thermalization noise of the mode $b_{s,cw}$, which can be canceled by applying a broadband squeezed-vacuum field \cite{PhysRevLett.120.093601,PhysRevA.100.062501,PhysRevLett.114.093602,PhysRevLett.128.083604} to reach the single-quantum level. Similarly, in Fig.\ref{fig.1}(b), the master equation is described as:

\begin{eqnarray}\label{eq9} 
	\frac{d \rho_2}{d t}&=&-i[\mathcal{H}_2,\rho_2]+\kappa_a L[a_{cw}]\rho_2+\kappa_b L[b_{ccw}]\rho_2\nonumber \\&&+\gamma_m(n_{\scriptstyle \mathrm{th}}+1)L[c]\rho_2+\gamma_m n_{\scriptstyle \mathrm{th}}L[c^{\dagger}]\rho_2,
\end{eqnarray}
where $\rho_2$ represents the density matrix of the system. It is worth noting that both types of photon detection channels are influenced by the environmental temperature $T$ due to the presence of the mechanical oscillator. In this work, we neglect the presence of thermal photons in the optical system, as their number is typically negligible compared to that of phonons. Therefore, the thermal bath is unlikely to have a significant impact on the results.

According to the input-output relation, we have $a_{\scriptstyle \mathrm{out}}=\varepsilon-\sqrt{\kappa_{ex1}}a$ and $b_{\scriptstyle \mathrm{out}}=\sqrt{\kappa_{ex2}}b$. The single photon transmissions are defined as

\begin{equation}\label{eq10} 
	T_{12/21}=\frac{\langle a_{\scriptstyle \mathrm{out}}^{\dagger}a_{\scriptstyle \mathrm{out}}\rangle}{\langle \varepsilon\varepsilon^*\rangle}, ~~~  T_{23}=\frac{\langle b_{\scriptstyle \mathrm{out}}^{\dagger}b_{\scriptstyle \mathrm{out}}\rangle}{\langle \varepsilon\varepsilon^{*}\rangle},
\end{equation}
where $\langle a_{\scriptstyle \mathrm{out}}^{\dagger}a_{\scriptstyle \mathrm{out}}\rangle$ and $\langle b_{\scriptstyle \mathrm{out}}^{\dagger}b_{\scriptstyle \mathrm{out}}\rangle$ are the mean photon numbers. $T_{12/21}$ represents the transmission from port 1 to port 2 or from port 2 to port 1, respectively, while $T_{23}$ is the transmission from port 2 to port 3. The corresponding photon isolation ratio in the lower bus waveguide is defined as $\eta = 10 \log_{10}(T_{21}/T_{12})$.

\section{Temperature sensitive nonreciprocal single photon transmission}\label{III}

\begin{figure}
	\centering
	\includegraphics[scale=0.85]{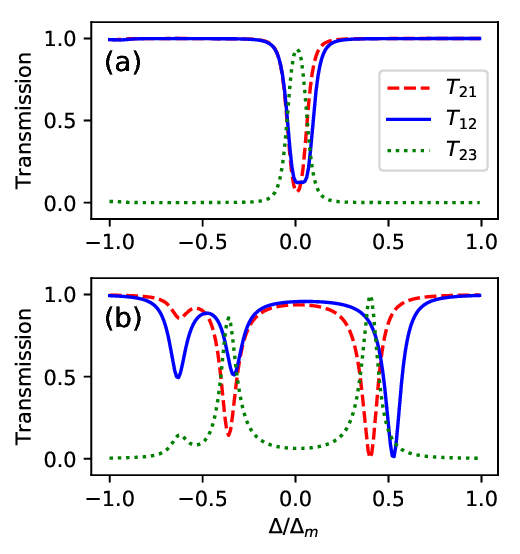}
	\caption{The transmittances $T_{12}$ (blue solid line), $T_{21}$ (red dashed line), and the transfer rate $T_{23}$ (green dotted line) as functions of $\Delta$ are shown. (a) corresponds to the scenario without NMS ($g_0=0.04\Delta_m$); (b) represents the scenario with NMS ($g_0=0.4\Delta_m$). Other parameters are: $\kappa=0.1\Delta_m$, $\gamma=0.007\Delta_m$, $\varepsilon=0.03\Delta_m$, $\Delta_s^{a/b}=0.01\Delta_m$, $J_0=0.09\Delta_m$, $\beta=0.9$, and $n_{\scriptstyle \mathrm{th}}=0$.}
	\label{fig.3}
\end{figure}

There are two different mechanisms for the coupling between two optical resonators \cite{PhysRevLett.128.083604}. When $g_0<\kappa$, the two resonators originally have no normal mode splitting (NMS). While for $g_0>\kappa$, the system switches to the NMS scenario. To accurately illustrate how the squeezed coupling strength $g_s$ influences the nonreciprocity of the system, assume the detuning $\Delta_s^b$ to be a small value. In Fig.\ref{fig.3}, we plot the single-photon transmission versus the detuning $\Delta$ for the no NMS scenario ($g_0=0.04\Delta_m$) and the NMS scenario ($g_0=0.4\Delta_m$), respectively. Since the value $\Delta_s^{a/b}=0.01\Delta_m$ is small enough, the mode resonance shift induced by quantum squeezing can be neglected.
In Fig.\ref{fig.3}(a), there is only one dip or peak in the single-photon transmission simulation curves, and no obvious nonreciprocity phenomenon is shown in the entire detuning range. On the contrary, for the NMS scenario in Fig.\ref{fig.3}(b), there are three main dips in both $T_{12}$ and $T_{21}$, indicating the system exhibits obvious single-photon nonreciprocal transmission characteristics, especially at the detuning $\Delta=g_{\nu}$ ($\nu=0,s$). Additionally, it is always approximately true that $T_{21}+T_{23}=1$. At the detuning $\Delta=g_0$, the single photon follows the transmission path $1\rightarrow 2\rightarrow 3$, indicating that the device operates as a three-port quasicirculator.

In this section, we explain the existence of dips in Fig.\ref{fig.3}(b). In the Hilbert space, $\vert n_a, n_b, n_c\rangle$ represents the Fock state with $n_a$ ($n_b$) photons in resonator $R_A$ ($R_B$) and $n_c$ phonons in resonator $C$. Supposing the initial state $\vert \Delta\rangle$ is in a subspace formed by $\lbrace\vert 1,0,0\rangle, \vert 1,0,1\rangle,\vert 1,0,2\rangle, \vert 0,1,0\rangle, \vert 0,1,1\rangle,\vert 0,1,2\rangle\rbrace$, where we truncate the subspace to a maximum phonon number of two. Then the matrix form of the Hamiltonian $\mathcal{H}_2$ without the probe field part reads:

\begin{equation}
\left( \begin{array}{cccccc}
\Delta&-J_0&0&g_0&0&0\\
-J_0&\Delta+\Delta_m&-\sqrt{2}J_0&0&g_0&0\\
0&-\sqrt{2}J_0&\Delta+2\Delta_m&0&0&g_0\\
g_0&0&0&\Delta&J_0&0\\
0&g_0&0&J_0&\Delta+\Delta_m&\sqrt{2}J_0\\
0&0&g_0&0&\sqrt{2}J_0&\Delta+2\Delta_m
\end{array} \right)
\end{equation}\label{eq11}

Solving the above Hamiltonian, we obtain partial solutions corresponding to three dips of $T_{21}$ in Fig.\ref{fig.3}(b), which are $\Delta_1=0.405\Delta_m$, $\Delta_2=-0.364\Delta_m$, and $\Delta_3=-0.627\Delta_m$. The eigenstate at $\Delta_1$ is expressed as:

\begin{eqnarray}\label{eq12}
	\vert \Delta_1\rangle &\simeq & \frac{1}{\mathcal{N}_1}[\alpha_0(\vert 1,0,0\rangle-\vert 0,1,0\rangle)+\alpha_1(\vert 1,0,1\rangle\nonumber \\&&+\vert 0,1,1\rangle)+\alpha_2(\vert 1,0,2\rangle-\vert 0,1,2\rangle)],
\end{eqnarray} 
where $\alpha_0=314\alpha_2$ and $\alpha_1=16\alpha_2$, with $\mathcal{N}_1$ representing the normalization constant.

Similarly, the matrix form of the Hamiltonian $\mathcal{H}_1^s$ without the probe field part reads:

\begin{equation}
	\left( \begin{array}{cccccc}
		\Delta&-J_0&0&g_s&0&0\\
		-J_0&\Delta+\Delta_m&-\sqrt{2}J_0&0&g_s&0\\
		0&-\sqrt{2}J_0&\Delta+2\Delta_m&0&0&g_s\\
		g_s&0&0&\Delta&J_s&0\\
		0&g_s&0&J_s&\Delta+\Delta_m&\sqrt{2}J_s\\
		0&0&g_s&0&\sqrt{2}J_s&\Delta+2\Delta_m\\
	\end{array} \right)
\end{equation}\label{eq13}

We also obtain partial solutions corresponding to three dips of $T_{12}$ in Fig.\ref{fig.3}(b), which are $\Delta_1^s=0.528\Delta_m$, $\Delta_2^s=-0.337\Delta_m$, and $\Delta_3^s=-0.635\Delta_m$. The eigenstate at $\Delta_1^s$ is expressed as:

\begin{eqnarray}\label{eq14} 
	\vert \Delta_1^s\rangle &\simeq & \frac{1}{\mathcal{N}_1^s}(\alpha_{01}^s\vert 1,0,0\rangle-\alpha_{02}^s\vert 0,1,0\rangle+\alpha_{11}^s\vert 1,0,1\rangle\nonumber \\&&+\alpha_{12}^s\vert 0,1,1\rangle+\alpha_{21}^s\vert 1,0,2\rangle-\alpha_{22}^s\vert 0,1,2\rangle),
\end{eqnarray}
where $\alpha_{01}^s=247\alpha_{21}$, $\alpha_{02}^s=253\alpha_{21}$, $\alpha_{11}^s=3\alpha_{21}$, $\alpha_{12}^s=34\alpha_{21}$, $\alpha_{22}^s=4\alpha_{21}$, with $\mathcal{N}_1^s$ representing the normalization constant. It is not difficult to see that the proportion of base vectors containing phonons is small for both cases.

\begin{figure}
	\centering
	\includegraphics[scale=0.8]{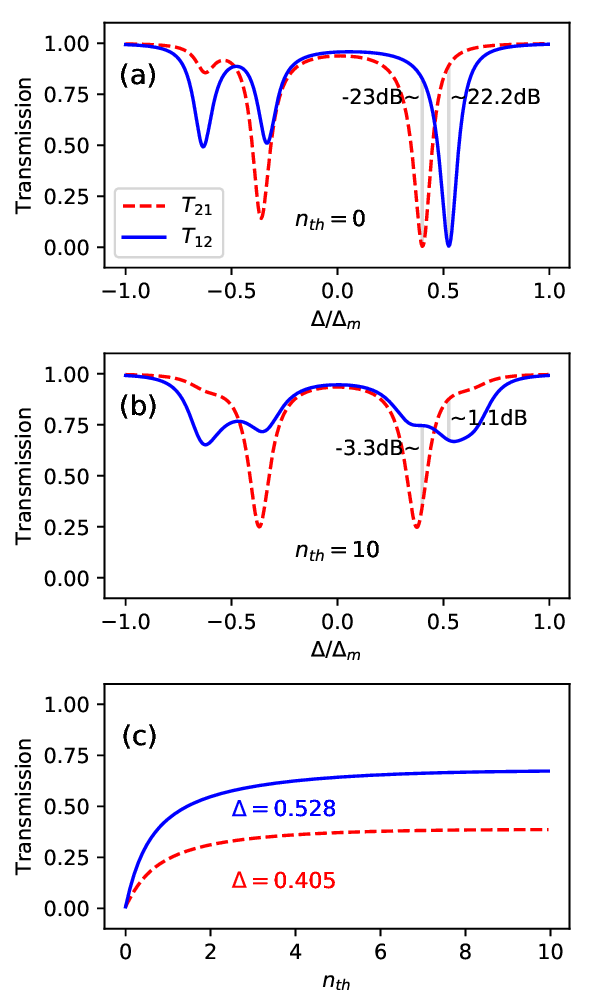}
	\caption{The transmittances $T_{12}$ (blue solid line) and $T_{21}$ (red dashed line) as functions of $\Delta$ for (a) $n_{\scriptstyle \mathrm{th}}=0$ and (b) $n_{\scriptstyle \mathrm{th}}=10$, respectively. In (c), the transmittances $T_{12}$ at detuning $\Delta=0.528 \Delta_m$ and the transmittances $T_{21}$ at detuning $\Delta=0.405 \Delta_m$ are plotted as functions of $n_{\scriptstyle \mathrm{th}}$. Other parameters are $\kappa=0.1\Delta_m$, $\gamma=0.007\Delta_m$, $\varepsilon=0.03\Delta_m$, $\Delta_s^{a/b}=0.01\Delta_m$, $g_0=0.4\Delta_m$, $J_0=0.09\Delta_m$, and $\beta=0.9$.}
	\label{fig.4}
\end{figure}

The system is thermalized by mechanical noise. In Fig.\ref{fig.4}(a) and Fig.\ref{fig.4}(b), we plot the single-photon transmission versus the detuning $\Delta$ for $n_{\scriptstyle \mathrm{th}}=0$ and $n_{\scriptstyle \mathrm{th}}=10$, respectively. In the first case, at detunings $\Delta_1$ and $\Delta_1^s$, the corresponding isolation ratios are $\eta=-23$ dB and $\eta=22.2$ dB, indicating strong nonreciprocity due to directional quantum squeezing in the low-$n_{\scriptstyle \mathrm{th}}$ regime.
However, as mechanical noise increases, the system's nonreciprocity is visibly impaired, with isolation ratios decreasing to $\eta = -3.3$ dB and $\eta = 1.1$ dB. These changes suggest that temperature significantly impacts the isolation effect of single photons.
To better illustrate the system's photon transport response to temperature variations, we plot the transmission $T_{21}$ at $\Delta_1$ and $T_{12}$ at $\Delta_1^s$ versus $n_{\scriptstyle \mathrm{th}}$ in Fig.\ref{fig.4}(c).
As \( n_{\mathrm{th}} \) increases, single-photon transmission initially rises and eventually stabilizes around \( n_{\mathrm{th}} = 10 \). This can be interpreted as follows: as the thermal phonon number increases, photon decoherence and scattering effects intensify, initially boosting the transmission probability. However, beyond a certain threshold, both the decoherence and scattering rates approach their saturation points. In other words, additional thermal noise no longer significantly increases coherence loss, and the photon scattering probability reaches a limit. As a result, the system's photonic transmission stabilizes, with the transmission probability becoming constant rather than continuing to vary with the thermal phonon number. Furthermore, directional parametric amplification enhances the optomechanical coupling strength \( J_s \), making the transmission \( T_{12} \) particularly sensitive to changes in \( n_{\mathrm{th}} \).

\begin{figure}
	\centering
	\includegraphics[scale=0.85]{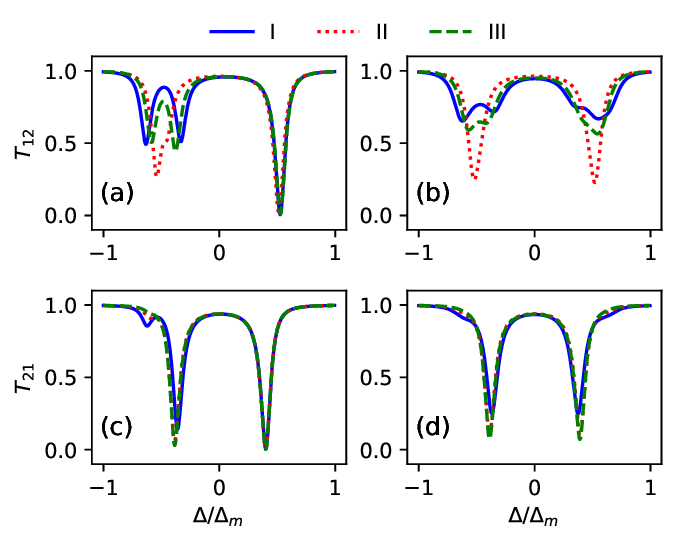}
	\caption{The transmittances $T_{12}$ in (a) and (b), and $T_{21}$ in (c) and (d) as functions of $\Delta$ for the case of the mechanical nanostring oscillator C being placed at area I (blue solid line), area II (red dotted line), and area III (green dashed line). In (a) and (c), $n_{\scriptstyle \mathrm{th}}=0$; in (b) and (d), $n_{\scriptstyle \mathrm{th}}=10$. Other parameters are $\kappa=0.1\Delta_m$, $\gamma=0.007\Delta_m$, $\varepsilon=0.03\Delta_m$, $\Delta_s^{a/b}=0.01\Delta_m$, $g_0=0.4\Delta_m$, $J_0=0.09\Delta_m$, and $\beta=0.9$.}
	\label{fig.5}
\end{figure}

In addition, the system’s sensitivity to temperature varies depending on the location of resonator C. We plot the single-photon transmission versus the detuning $\Delta$ for $n_{\scriptstyle \mathrm{th}}=0$ in Fig.\ref{fig.5}(a)(c) and for $n_{\scriptstyle \mathrm{th}}=10$ in Fig.\ref{fig.5}(b)(d). The blue solid line represents resonator C placed in area I, the red dotted line represents it in area II, and the green dashed line represents it in area III.
Comparing Fig.\ref{fig.5}(a) and Fig.\ref{fig.5}(b), we observe that resonator C in area I shows the strongest response. This is because photons in both the $a$-mode and $b$-mode are converted into phonons; a higher phonon conversion rate increases the system's sensitivity to temperature. In contrast, resonator C in area II shows a more pronounced response than in area III, as the directional parametric amplification enhances the interaction between $b$-mode photons and phonons.
Fig.\ref{fig.5}(c)(d) illustrates the case when the photon incidence direction is reversed. Here, the pump field does not interfere with photon transmission, and the position of resonator C has minimal impact on $T_{21}$, even with temperature changes.

\section{Temperature sensitive nonreciprocal photon correlations}\label{IV}

In this work, we use a coherent probe beam to drive the $\chi^{(2)}$ nonlinear microring systems \cite{PhysRevLett.121.153601}, allowing us to study the response of the system's photon statistical properties to temperature. By applying the input-output relations, the statistical properties of the transmitted photons can be described by the second-order correlation functions as follows:

\begin{equation}\label{eq15} 
	g_{12/21}^{(2)}(0)=\frac{\langle a_{\scriptstyle \mathrm{out}}^{\dagger 2}a_{\scriptstyle \mathrm{out}}^2\rangle}{\langle  a_{\scriptstyle \mathrm{out}}^{\dagger}a_{\scriptstyle \mathrm{out}}\rangle^2},  
\end{equation}
where $g_{12/21}^{(2)}(0)$ is the equal-time second-order correlation function for the output photons from port 1 or port 2.

Based on the calculations in Section \ref{III}, we find that
the eigenvectors with zero phonon number constitute a
significant portion of the eigenstates, including $\vert\Delta_1 \rangle$ and $\vert\Delta_1^s \rangle$. To approximate the two-photon resonance, we select the subspace $\lbrace \vert 2,0 \rangle, \vert 1,1 \rangle , \vert 0,2 \rangle\rbrace$. The matrix forms of the Hamiltonians  $\mathcal{H}_2$ and $\mathcal{H}_1^s$ without the probe field
component are then given by

\begin{equation}\label{eq16} 
	\left( \begin{array}{ccc}
		2\Delta&\sqrt{2}g_\nu&0\\
		\sqrt{2}g_\nu&2\Delta&\sqrt{2}g_\nu\\
		0&\sqrt{2}g_\nu&2\Delta\\
	\end{array} \right)
\end{equation}
we obtain the solutions $\Delta_{\pm\nu}=\pm g_{\nu}$ and $\Delta^{'}=0$. The
corresponding eigenstates are expressed as

\begin{equation}\label{eq17} 
	\vert \Delta_{\pm\nu} \rangle=\frac{1}{\mathcal{N}_2}(\vert 2,0 \rangle\pm\sqrt{2}\vert 1,1 \rangle+\vert 0,2 \rangle),
\end{equation}

\begin{equation}
	\vert \Delta^{'} \rangle=\frac{1}{\mathcal{N}_3}(\vert 2,0 \rangle+\vert 0,2 \rangle),
\end{equation}\label{eq18} 
where $\mathcal{N}_2$ and $\mathcal{N}_3$ are normalization constants.  It’s evident that $\Delta_\nu$ represents both single and two-photon resonances of the system, making it an optimal point for detecting the statistical properties of photons in this quantum circulator. Conversely, in the case of $\Delta_{-\nu}$ and $\Delta^{'}$, the detection port exhibits both single-photon and two photon characteristics, resembling a coherent state.

\begin{figure}
	\includegraphics[scale=0.8]{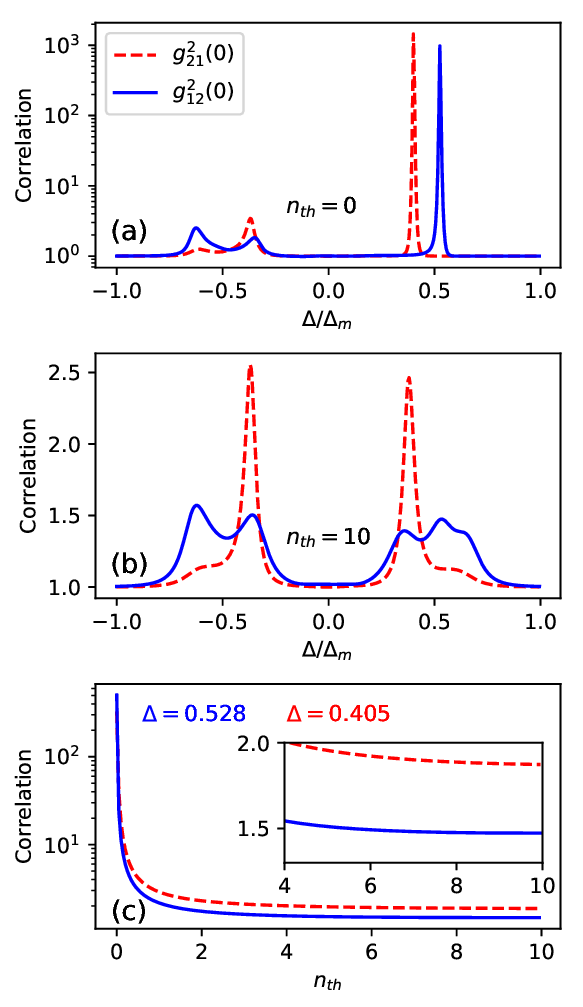}
	\centering
	\caption{The equal time second-order correlation functions $g_{12}^2(0)$ (blue solid line) and $g_{21}^2(0)$ (red dashed line) as functions of $\Delta$ for (a) $n_{\scriptstyle \mathrm{th}}=0$ and (b) $n_{\scriptstyle \mathrm{th}}=10$, respectively. In (c), $g_{12}^2(0)$ at detuning $\Delta=0.528 \Delta_m$ and $g_{21}^2(0)$ at detuning $\Delta=0.405 \Delta_m$ are plotted as functions of $n_{\scriptstyle \mathrm{th}}$. Other parameters are $\kappa=0.1\Delta_m$, $\gamma=0.007\Delta_m$, $\varepsilon=0.03\Delta_m$, $\Delta_s^{a/b}=0.01\Delta_m$, $g_0=0.4\Delta_m$, $J_0=0.09\Delta_m$, and $\beta=0.9$.}
	\label{fig.6}
\end{figure}

\begin{figure}
	\centering
	\includegraphics[scale=0.85]{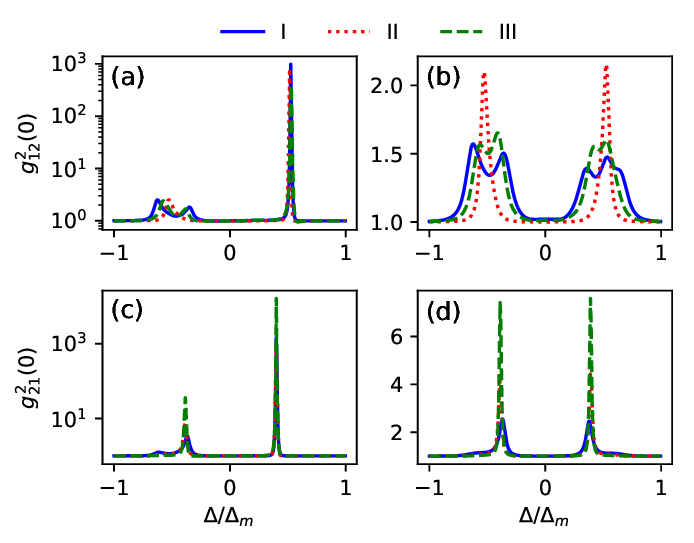}
	\caption{The equal-time second-order correlation functions $g_{12}^2(0)$ in (a) and (b) and $g_{21}^2(0)$ in (c) and (d) as functions of $\Delta$ for the mechanical nanostring oscillator C placed in area I (blue solid line), area II (red dotted line), and area III (green dashed line). In (a) and (c), $n_{\scriptstyle \mathrm{th}}=0$; in (b) and (d), $n_{\scriptstyle \mathrm{th}}=10$. Other parameters are $\kappa=0.1\Delta_m$, $\gamma=0.007\Delta_m$, $\varepsilon=0.03\Delta_m$, $\Delta_s^{a/b}=0.01\Delta_m$, $g_0=0.4\Delta_m$, $J_0=0.09\Delta_m$, and $\beta=0.9$.}
	\label{fig.7}
\end{figure}

In Fig.\ref{fig.6}, we depict the second-order correlation functions against the detuning $\Delta$ for $n_{\scriptstyle \mathrm{th}}=0$ in Fig.\ref{fig.6}(a) and $n_{\scriptstyle \mathrm{th}}=10$ in Fig.\ref{fig.6}(b), respectively. In the former case, numerical simulations reveal that $g_{21}^{(2)}(0)\gg 1$ and $g_{12}^{(2)}(0)\approx 1$ at $\Delta=\Delta_0$, while $g_{12}^{(2)}(0)\gg 1$ and $g_{21}^{(2)}(0)\approx 1$ at $\Delta=\Delta_s$. The transmitted photons exhibit a nonreciprocal strong bunching effect with low transmissibility since the single photons are transmitted to the upper drop waveguide  ports. However, when $n_{\scriptstyle \mathrm{th}}=10$, the statistical properties of the photons undergo significant changes, manifesting as a weak bunching effect. The physical explanation is that thermal phonons introduce phase and amplitude noise, enhance multi-phonon scattering, and induce decoherence, thereby disrupting the nonclassical statistical properties of the optical field. Consequently, the photon bunching effect is diminished, and the photon statistics tend to approach a Poisson distribution.
In Fig.\ref{fig.6}(c), we illustrate the second-order correlation functions $g_{12}^{(2)}(0)$ at detuning $\Delta=0.528\Delta_m$ and $g_{21}^{(2)}(0)$ at detuning $\Delta=0.405\Delta_m$ against the mean thermal phonon number $n_{\scriptstyle \mathrm{th}}$. We observe that the correlation functions decrease exponentially before $n_{\scriptstyle \mathrm{th}}=1$ and then remain constant near $n_{\scriptstyle \mathrm{th}}=10$. Remarkably, the statistical properties of photons transmitted in the opposite directions exhibit no significant differences with changes in temperature, indicating that the directional quantum squeezing effect on improving photon statistics due to temperature is not apparent. This phenomenon contrasts with single-photon transmission.
Furthermore, we examine the correlation functions corresponding to the location of resonator C in Fig.\ref{fig.7}. Comparing Fig.\ref{fig.7}(a) and Fig.\ref{fig.7}(b) at detuning $\Delta=0.528\Delta_m$, and Fig.\ref{fig.7}(c) and Fig.\ref{fig.7}(d) at detuning $\Delta=0.405\Delta_m$, we observe that the different locations of resonator C do not distinctly distinguish the statistical properties of photons, both exhibiting a transition from strong bunching effect at $n_{\scriptstyle \mathrm{th}}=0$ to weak bunching effect at $n_{\scriptstyle \mathrm{th}}=10$.

\begin{figure}
	\centering
	\includegraphics[scale=0.8]{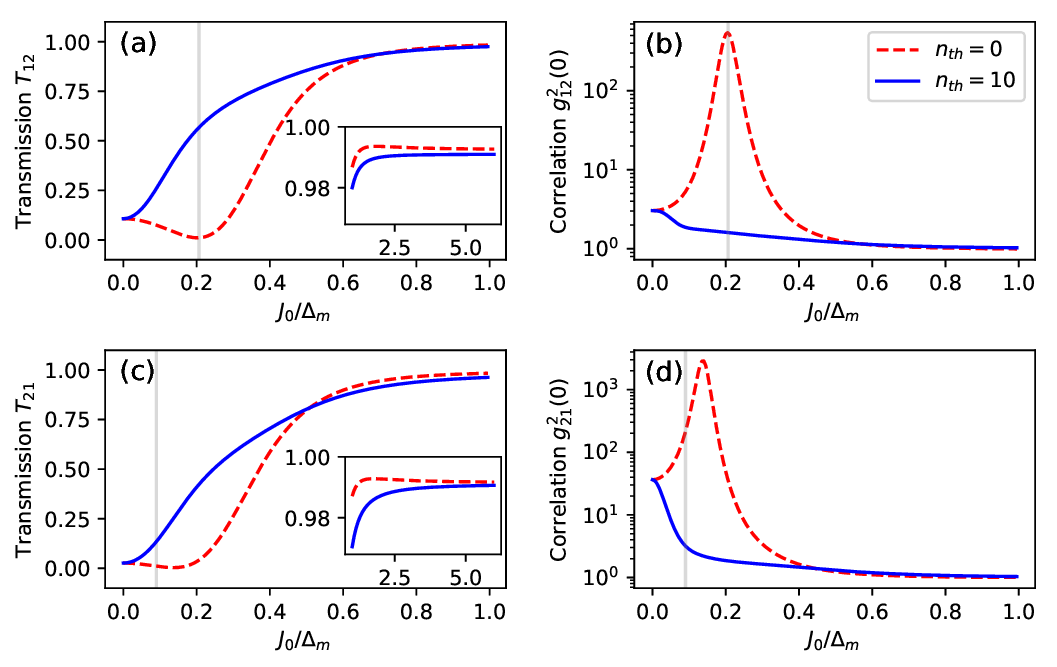}
	\caption{The transmittances $T_{12}$ in (a) and $T_{21}$ in (b) as functions of the resonator-oscillator coupling strength $J_0$, with the oscillator C placed in area III. The equal-time second-order correlation function $g_{12}^2(0)$ in (c) and $g_{21}^2(0)$ in (d) as functions of $J_0$. The red dashed line represents $n_{\scriptstyle \mathrm{th}}=0$ and the blue solid line represents $n_{\scriptstyle \mathrm{th}}=10$. The detuning in (a) and (b) is $\Delta=0.528\Delta_m$, while in (c) and (d) it is $\Delta=0.405\Delta_m$. Other parameters are $\kappa=0.1\Delta_m$, $\gamma=0.007\Delta_m$, $\varepsilon=0.03\Delta_m$, $g_0=0.4\Delta_m$, and $\beta=0.9$.}
	\label{fig.8}
\end{figure}

We further investigate the influence of the resonator-oscillator coupling strength $J_0$ on the nonreciprocity of the temperature-dependent system. In Fig.\ref{fig.8}, we illustrate the single-photon transmission and correlation functions against $J_0$ for the cases of $n_{\scriptstyle \mathrm{th}}=0$ and $n_{\scriptstyle \mathrm{th}}=10$, respectively. In Fig.\ref{fig.8}(a) and Fig.\ref{fig.8}(c), for $n_{\scriptstyle \mathrm{th}}=10$, the value of $T_{12/21}$ continuously increases until it approaches $J_0=1\Delta_m$. Conversely, for $n_{\scriptstyle \mathrm{th}}=0$, the value of $T_{12/21}$ initially decreases before continuously increasing until it approaches $J_0=1\Delta_m$. This phenomenon can be explained by the fact that as $J_0$ increases, the exchange of photons and thermal phonons also increases, dissipating onto the mechanical oscillator. However, as $J_0$ further increases, oscillator C acts as a scattering center, re-scattering the $b$-mode photons originally emitted from port 4 or port 3 back into the resonant $R_A$, ultimately outputting to port 2 or port 1. The evolutionary behavior of $g_{12/21}^{(2)}(0)$ in Fig.\ref{fig.8}(b) and Fig.\ref{fig.8}(d) corresponds to the single-photon transmission case discussed earlier.
Additionally, analyzing Fig.\ref{fig.8}(a) and Fig.\ref{fig.8}(b), we find that the suitable coupling strength arises at around $J_0=0.09\Delta_m$, where $T_{12}$ and $g_{12}^{(2)}(0)$ are most responsive to temperature simultaneously. The corresponding effective coupling strength is $J_s=0.206\Delta_m$. However, for $T_{21}$ and $g_{21}^{(2)}(0)$ in Fig.\ref{fig.8}(c) and Fig.\ref{fig.8}(d), the most responsive point to temperature is not achieved, since the coupling strength for this case is not enhanced.

\section{Experimental feasibility}\label{V}
The thin-film material lithium niobate \cite{bruch201817,zhang2019broadband,PhysRevApplied.16.064004} has been extensively utilized to fabricate $\chi^{(2)}$ nonlinearity resonators. Recent advancements have led to the development of high-Q whispering gallery resonators on-chip, reaching Q-factors as high as $3\times 10^6$.
Additionally, epitaxial aluminum nitride has emerged as a promising photonic platform, showcasing efficient $\chi^{(2)}$ and $\chi^{(3)}$ nonlinear processes \cite{liu2019beyond,liu2017integrated}. It stands as a strong alternative to traditional lithium niobate materials.
Silicon has enabled the creation of ring resonators with unprecedentedly small sizes. Microring resonators and waveguides made of silicon are typically fabricated using techniques like e-beam or optical lithography and reactive ion etching, often employing CMOS manufacturing tools \cite{mcnab2003ultra,vlasov2004losses}.
In our proposed scheme, the prevalent method of coupling light from a waveguide to a ring resonator or between ring resonators is through a directional coupler \cite{bogaerts2012silicon}.

The whispering-gallery-mode optomechanical system enables the evanescent coupling of nanomechanical oscillators to an optical microresonator. This has been demonstrated in various setups, such as a tapered-fibre-interfaced optical microresonator dispersively coupled to an array of nanomechanical oscillators. For instance, doubly clamped SiN nanostring oscillators with dimensions of 110 nm $\times$ (300-500) nm $\times$ (15-40) $\mu$m have been utilized for this purpose \cite{anetsberger2009near}. Additionally, a high-stress $\textrm{SI}_3\textrm{N}_4$  nanomechanical beam integrated into the evanescent mode volume of a $\textrm{SIO}_2$  microdisk has been employed \cite{PhysRevApplied.5.054019}.
Apart from the traditional minimal model (i.e., one optical mode coupled to one mechanical mode) of cavity optomechanics, it is theoretically feasible to extend to multimode optomechanics, as discussed in Ref. \cite{RevModPhys.86.1391}.

Broadband squeezed-vacuum fields have proven to be an effective strategy for mitigating quantum noise, and they are currently utilized in advanced detectors. In recent experiments, a suspended 300-meter-long filter cavity capable of inducing a rotation of the squeezing ellipse below 100 Hz has demonstrated the reduction of quantum noise in advanced gravitational-wave detectors across their entire observation bandwidth \cite{PhysRevLett.124.171101}. Another study achieved broadband reduction of quantum radiation pressure noise \cite{yap2020broadband}.
In addition to these applications, broadband squeezed-vacuum fields have been employed in various physics experiments, including the suppression of radiative decay of atoms \cite{PhysRevLett.93.023005}, squeezing of mechanical modes in optomechanical systems \cite{PhysRevA.88.013835}, and the creation of squeezed lasing \cite{PhysRevLett.127.183603}, among others.

\section{Conclusion}\label{VI}
We have successfully demonstrated nonreciprocal single-photon transmission and quantum correlations in temperature-sensitive optomechanical systems, where the nonreciprocity arises from directional quantum squeezing. Our findings reveal that the nonreciprocity of the system is significantly influenced by thermal phonons.
Specifically, at detuning $\Delta=0.528\Delta_m$ (or $\Delta=0.405\Delta_m$), when $n_{\scriptstyle \mathrm{th}}=0$, the configuration behaves like a diode with a high isolation ratio, and the statistical properties of the transmitted photons exhibit a strong bunching effect. Conversely, for $n_{\scriptstyle \mathrm{th}}=10$, the isolation of the system is greatly reduced from 22.2 dB to 1.1 dB (or from -23 dB to -3.3 dB), and the statistical properties of the transmitted photons exhibit a weak bunching effect.
Moreover,  our setup benefits from directional parametric amplification, which enhances the optomechanical coupling strength. This enhancement makes the system more sensitive to temperature, making it suitable for precise temperature measurements at ultralow temperatures. For instance, we can use the values of the single-photon transmission rate or the equal-time second-order correlation function at detuning $\Delta_1$ and detuning $\Delta_1^s$ as temperature indicators.
In conclusion, our research provides a scheme for achieving optical nonreciprocity in chip-compatible magnetic-free devices, enabling backaction-immune quantum measurement and chiral quantum photonics. Furthermore, our work has the potential to be extended to phonon systems, allowing for applications such as phonon blockades \cite{PhysRevB.92.115407,PhysRevApplied.17.054004} and nonreciprocal phonon lasers \cite{PhysRevLett.104.083901,PhysRevLett.113.053604}.

\section*{Acknowledgments}
This work is supported by the National Natural Science
Foundation of China under Grant No. 12375018.

\section*{Data availability statement}
The data that support the findings of this study are available upon reasonable request from the authors.

\appendix
\renewcommand{\theequation}{A\arabic{equation}}
\setcounter{equation}{0}

\section*{Appendix A: Optomechanically coupled to two optical resonators with coupling strengths of opposite signs}\label{AppendixA}

\renewcommand{\thefigure}{A}
\begin{figure}
	\centering
	\includegraphics[scale=1]{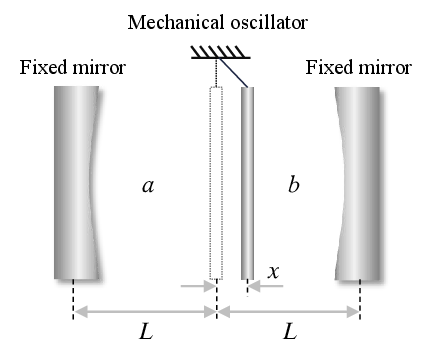}
	\caption{Equivalent schematic diagram of Fig. \ref{fig.1}. The optical cavity is divided into cavity $a$ and cavity $b$ by the central mechanical oscillator. At rest, the distance between the mechanical oscillator and the fixed mirrors at both ends is $L$.}
	\label{fig.A}
\end{figure}

The equivalent schematic diagram of Fig. \ref{fig.1} is shown in Fig. \ref{fig.A}. It is worth noting that the walls of the microring resonators in Fig. \ref{fig.1} are fixed (which contrasts with the configuration in Ref. \cite{agasti2023back}, where the system can be stretched or compressed), meaning the frequencies of the optical modes $\omega_a$ and $\omega_b$ remain unchanged. However, we have demonstrated that the configuration in Fig. \ref{fig.1} is mathematically equivalent to the configuration in Fig. \ref{fig.A}.

The resonant frequency of the cavity field is closely related to the cavity length. When the mechanical oscillator vibrates near its equilibrium position, the corresponding cavity length becomes $L\rightarrow L\pm x$, causing a shift in the cavity mode frequency \cite{meystre2017cavity}.

\begin{eqnarray}\label{eqA1} 
\omega_{a} (L+x)=\frac{\pi c}{L+x}, \nonumber\\
\omega_{b} (L-x)=\frac{\pi c}{L-x},
\end{eqnarray}
where $c$ denotes the speed of light in a vacuum.

In general, the displacement of the macroscopic oscillator is much smaller than the length of the optical cavity at rest, i.e., $x \ll L$, as described by the Taylor expansion formula
\begin{eqnarray}\label{eqA2} 
\omega_{a} &(L+x)&=\frac{\pi c}{L}\times\frac{1}{1+x/L}=\frac{\pi c}{L}\left(1-\frac{x}{L}+\frac{x^2}{L^2}-...\right)\nonumber\\
& & \approx\omega_{a} (L)\left(1-\frac{x}{L}\right),\nonumber\\
\omega_{b} &(L-x)&=\frac{\pi c}{L}\times\frac{1}{1-x/L}=\frac{\pi c}{L}\left(1+\frac{x}{L}+\frac{x^2}{L^2}+...\right)\nonumber\\
& &\approx\omega_{b} (L)\left(1+\frac{x}{L}\right).
\end{eqnarray}

When the optical field is driven by an external laser, the photons couple with the oscillator, and the total Hamiltonian of the system is given by

\begin{eqnarray}\label{eqA3} 
H_A+H_B+&H_{int}&=\omega_a(L+x)a^{\dagger}a+\omega_b(L-x)b^{\dagger}b\nonumber\\
& &=\omega_a(L)a^{\dagger}a+\omega_b(L)b^{\dagger}b\nonumber\\
& &~~~-\omega_a(L)\frac{x}{L}a^{\dagger}a+\omega_b(L)\frac{x}{L}b^{\dagger}b.
\end{eqnarray}
The equation indicates that the free Hamiltonian of the optical modes $H_A$ and $H_B$ can be considered constant, corresponding to the scenario shown in Fig. \ref{fig.1}. Substituting $x=\sqrt{\frac{1}{2 m \omega_m}}(c+c^{\dagger})$ (where $m$ is the effective mass of the mechanical oscillator) into the equation and assuming $\omega_a(L)=\omega_b(L)=\omega(L)$, we obtain:

\begin{equation}\label{eqA4} 
H_{int}=-\frac{\omega(L)}{L}(a^{\dagger}a-b^{\dagger}b)x=-J_0^{'}(a^{\dagger}a-b^{\dagger}b)(c+c^{\dagger}),
\end{equation}
where $J_0^{'}=\sqrt{\frac{1}{2 m \omega_m}}\frac{\omega(L)}{L}$.

\renewcommand{\theequation}{B\arabic{equation}}
\setcounter{equation}{0}

\section*{Appendix B: Detailed solution of Eq. \ref{eq6}}\label{AppendixB}

\setcounter{equation}{0}

When transforming the system into the squeezing picture, we should only substitute $b_{cw}$ ($b_{cw}^{\dagger}$) in Eq.\ref{eq5}. Thus, the Hamiltonian we need to consider simplifies to

\begin{eqnarray}\label{eqB1} 
\mathcal{H}&=&\Delta_s^b b^{\dagger}_{cw}b_{cw}+\frac{\Omega}{2}(b^{\dagger 2}_{cw}+b^2_{cw})+g_0(a^{\dagger}_{ccw}b_{cw}+a_{ccw}b^{\dagger}_{cw})\nonumber\\
& &+J_0 b^{\dagger}_{cw}b_{cw}(c^{\dagger}+c).
\end{eqnarray}
By applying the Bogoliubov squeezing transformation $b_{cw} = \cosh(r_s) b_{s,cw} - \sinh(r_s) b_{s,cw}^\dagger$ and the commutation relation $[b_{s,cw}, b_{s,cw}^{\dagger}]=1$ to Eq. \ref{eqB1}, neglecting the parametric interaction term $J_0 \textrm{sinh}(2 r_s)(b^{\dagger 2}_{s,cw}+b_{s,cw}^2)(c^{\dagger}+c)/2$,  and under the rotating-wave approximation while neglecting the counter-rotating terms, the Hamiltonian becomes

\begin{eqnarray}\label{eqB2} 
\mathcal{H}^s&=&[\Delta_s^b\textrm{cosh}(2 r_s)-\Omega \textrm{sinh}(2 r_s)]
b^{\dagger}_{s,cw}b_{s,cw}\nonumber \\&&+[\frac{\Omega}{2}\textrm{cosh}(2 r_s)-\frac{\Delta_s^b}{2}\textrm{sinh}(2 r_s)](	b^{\dagger 2}_{s,cw}+b_{s,cw}^2)\nonumber \\&&+[J_s b^{\dagger}_{s,cw}b_{s,cw}+J_0 \textrm{sinh}^2(r_s)](c^{\dagger}+c)\nonumber \\&&+g_s(a^{\dagger}_{ccw}b_{s,cw}+a_{ccw}b^{\dagger}_{s,cw})\nonumber \\&&+\Delta_s^b \textrm{sinh}^2(r_s)-\frac{\Omega}{2}\textrm{sinh}(2 r_s),
\end{eqnarray}
neglecting the last constant term, and in the frame rotating at the frequency $\Delta_p$, the above equation becomes

\begin{eqnarray}\label{eqB3} 
\mathcal{H}^s&=&\Delta_b^s b^{\dagger}_{s,cw}b_{s,cw}+g_s(a^{\dagger}_{ccw}b_{s,cw}+a_{ccw}b^{\dagger}_{s,cw})\nonumber \\&&+[J_s b^{\dagger}_{s,cw}b_{s,cw}+J_0 \textrm{sinh}^2(r_s)](c^{\dagger}+c).
\end{eqnarray}
The last term, which applies to the mechanical oscillator, can be canceled by applying a constant force $F$. The remaining unchanged terms from Eq. \ref{eq5} are then incorporated, leading to the final form presented in Eq. \ref{eq6}.

\renewcommand{\theequation}{C\arabic{equation}}
\setcounter{equation}{0}

\section*{Appendix C: The effect of the omitted optomechanical coupling terms on photon transmission}\label{AppendixC}

\setcounter{equation}{0}

\renewcommand{\thefigure}{C1}
\begin{figure}
	\centering
	\includegraphics[scale=0.8]{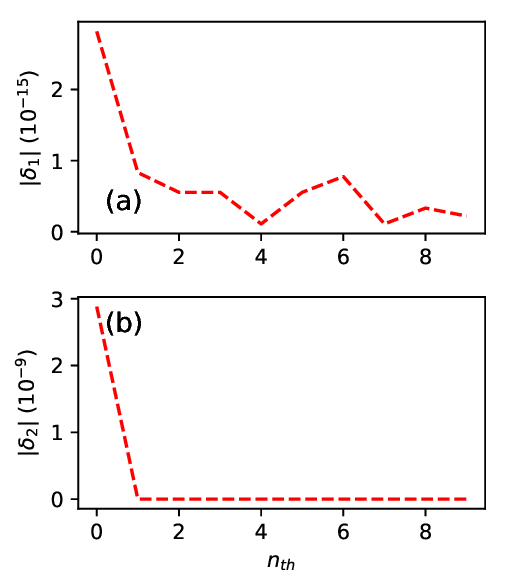}
	\caption{The parameters \( |\delta_1| \) and \( |\delta_2| \) as functions of \( n_{th} \) at a detuning of \( \Delta = 0.528 \Delta_m\). The other parameters are \( \kappa = 0.1 \Delta_m \), \( \gamma = 0.007 \Delta_m \), \( \varepsilon = 0.03 \Delta_m \), \( \Delta_s^{a/b} = 0.01 \Delta_m \), \( g_0 = 0.4 \Delta_m \), \( J_0 = 0.09 \Delta_m \), and \( \beta = 0.9 \).
		}
	\label{fig.C1}
\end{figure}

In this section, we incorporate the coupling of oscillator \( C \) with the CCW mode in \( R_B \) and the CW mode in \( R_A \) into the system when a photon is incident from port 1. The corresponding Hamiltonian in the squeezing picture is given by:

\begin{eqnarray}\label{eqC1} 
(\mathcal{H}^{s}_1)^{'}&=&\mathcal{H}^s_1+\Delta_a a^{\dagger}_{cw}a_{cw}+\Delta_b b^{\dagger}_{ccw}b_{ccw}+g_0(a^{\dagger}_{cw}b_{ccw}+a_{cw}b^{\dagger}_{ccw})\nonumber \\&&
-(J_0 a^{\dagger}_{cw}a_{cw}-J_0 b^{\dagger}_{ccw}b_{ccw})(c^{\dagger}+c).
\end{eqnarray}
Similar improvements need to be made to the master equation, where we must account for the dissipation of photons in these two modes as:

\begin{eqnarray}\label{eqC2} 
\frac{d \rho_1^{'}}{d t}&=&-i[(\mathcal{H}^{s}_1)^{'},\rho_1^{'}]+\kappa_a L[a_{ccw}]\rho_1^{'}+\kappa_a L[a_{cw}]\rho_1^{'}+\kappa_b L[b_{s,cw}]\rho_1^{'}+\kappa_b L[b_{ccw}]\rho_1^{'}\nonumber \\&&+\gamma_m(n_{\scriptstyle \mathrm{th}}+1)L[c]\rho_1^{'}+\gamma_m n_{\scriptstyle \mathrm{th}}L[c^{\dagger}]\rho_1^{'}\nonumber \\&&+\mathcal{L}_n[b_{s,cw}]\rho_1^{'}.
\end{eqnarray}
From the complete Hamiltonian and master equation above, we obtain the single-photon transmission \( T_{12}' \) and the second-order correlation function \( g_{12}'(0) \). We then define \( \delta_1 = T_{12} - T_{12}' \) and \( \delta_2 = g_{12}(0) - g_{12}'(0) \) to evaluate the difference in photon transmission between the two systems. In Fig. \ref{fig.C1}, we plot the parameters \( |\delta_1| \) and \( |\delta_2| \) as functions of \( n_{th} \) at a detuning of \( \Delta = 0.528 \Delta_m \). It is evident that these two parameters are of the order of \( 10^{-15} \) and \( 10^{-9} \), respectively, indicating that there is little difference between the two systems in terms of single-photon transmission and photon statistical properties. Therefore, the two optomechanical coupling terms omitted in the main text have a minimal impact on our observed photon transmission effects and can be safely neglected.

\renewcommand{\thefigure}{C2}
\begin{figure}
	\centering
	\includegraphics[scale=0.8]{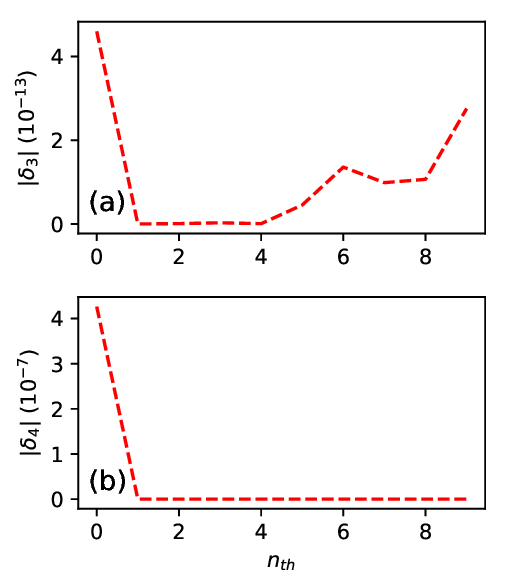}
	\caption{The parameters \( |\delta_3| \) and \( |\delta_4| \) as functions of \( n_{th} \) at a detuning of \( \Delta = 0.405 \Delta_m\). The other parameters are \( \kappa = 0.1 \Delta_m \), \( \gamma = 0.007 \Delta_m \), \( \varepsilon = 0.03 \Delta_m \), \( \Delta_s^{a/b} = 0.01 \Delta_m \), \( g_0 = 0.4 \Delta_m \), \( J_0 = 0.09 \Delta_m \), and \( \beta = 0.9 \).
	}
	\label{fig.C2}
\end{figure}

Similarly, when a photon is incident from port 2, with an additonal force $F$  applied to  oscillator \( C \), the corresponding Hamiltonian in the squeezing picture is given by:

\begin{eqnarray}\label{eqC3} 
	(\mathcal{H}^{s}_2)^{'}&=&\mathcal{H}_2+\Delta_a a^{\dagger}_{ccw}a_{ccw}+\Delta_b^s b^{\dagger}_{cw}b_{cw}+g_s(a^{\dagger}_{ccw}b_{cw}+a_{ccw}b^{\dagger}_{cw})\nonumber \\&&
	-(J_0 a^{\dagger}_{ccw}a_{ccw}-J_s b^{\dagger}_{cw}b_{cw})(c^{\dagger}+c).
\end{eqnarray}
For the master equation, it is sufficient to replace $(\mathcal{H}^{s}_1)^{'}$ with $ (\mathcal{H}^{s}_2)^{'}$ and $\rho_1^{'}$ with $ \rho_2^{'}$ in Eq. \ref{eqC2}. Based on these, we obtain the single-photon transmission \( T_{21}' \) and the second-order correlation function \( g_{21}'(0) \). We then define \( \delta_3 = T_{21} - T_{21}' \) and \( \delta_4 = g_{21}(0) - g_{21}'(0) \) to evaluate the difference in photon transmission between the two systems as before. In Fig. \ref{fig.C2}, we plot the parameters \( |\delta_3| \) and \( |\delta_4| \) as functions of \( n_{th} \) at a detuning of \( \Delta = 0.405 \Delta_m \). The simulation results indicate that these two parameters are on the order of \( 10^{-13} \) and \( 10^{-7} \), respectively. Consequently, the corresponding optomechanical coupling terms can be safely neglected.

\setcounter{section}{0} 
\renewcommand{\thesection}{\arabic{section}} 
\section*{References}

\providecommand{\newblock}{}

\end{document}